\documentclass[aps,prd,superscriptaddress,showpacs, onecolumn,12pt]{revtex4-1}
\usepackage{graphicx}
\usepackage{epstopdf}
\usepackage{amsmath}
\usepackage{amsfonts}
\usepackage{amssymb}
\usepackage{latexsym}
\usepackage{hyperref}
\usepackage[english]{babel}
\usepackage[utf8]{inputenc}
\usepackage[colorinlistoftodos]{todonotes}
\usepackage{xcolor}
\usepackage{slashed} 
\usepackage{bm}  
\begin{document}
\title{Fermionic Solutions of a Five-Dimensional Chern-Simons AdS Supergravity without Gravitino}
\author{Y.M.P. Gomes}\email{ymuller@cbpf.br}
\affiliation{Centro Brasileiro de Pesquisas F\'{i}sicas (CBPF), Rua Dr Xavier Sigaud 150, Urca, Rio de Janeiro, Brazil, CEP 22290-180}
\begin{abstract}
\paragraph*{}Based on recent discussions on the so-called unconventional supersymmetry, we analyze a class of solutions of a 5D Chern-Simons AdS-$\mathcal{N}$-SUGRA formulation without gravitino fields.  With a Randall-Sundrum-type ansatz, we exploit the properties of Chern-Simons theories to find solutions to the fermionic field equations in a particular dimensional reduction context. We show that this specific dimensional reduction yields a non-trivial equation of motion for the fermionic field. We actually get a non-linear equation of motion for the fermionic fields, typical of models where torsion is present. This fermionic equation describes massive fermions with specific couplings with the bosonic supersymmetric degrees of freedom and we show that, in some specific limits, we can infer about the localization of the fermions' chirality components by means of the particular function that comes out to the dimensional reduction scheme.
\end{abstract} 
\pacs{04.65.+e,11.15.Yc}
\maketitle
\section{introduction} 
\paragraph*{}An alternative method to build up a theory with supersymmetry (SUSY) is by implementing a gauge theory for a super-algebra that includes an internal gauge group, $\mathcal{G}$, along with a local $SO(1,D-1)$ algebra that has to be set up to connect these two symmetries through fermionic supercharges \cite{uncon,uncon1,uncon2}. In these references, the field multiplet is composed by a (non-)Abelian field, $A$, a spin-1/2 Dirac fermion, $\psi$, the spin connection, $\omega^{ab}$, the $d$-bien, $e^a$, and additional gauge fields which  complete the degrees of freedom to accomplish the supersymmetrization. These additional fields are dependent on the structure of the group and the space-time which we intend to work in. The representations of the fields are not all the same. The Dirac spinor transforms under the fundamental representation, while the gauge connection belongs to the adjoint representation of $\mathcal{G}$. In this framework, the metric is completely invariant under the symmetries $\mathcal{G}$, $SO(1,D-1)$ and supersymmetry.
\paragraph*{}Due to the properties quoted above, the model displays important differences in comparison with standard SUSYs. For example, there is no superpartners with degenerate masses, nor an equal number of degrees of freedom of bosons and fermions. There is not even a spin-3/2 fermion, i.e., a gravitino, in the spectrum of the model \cite{uncon,uncon1,uncon2}.
\paragraph*{} It is remarkable that, in odd dimensions, the Chern-Simons (CS) form is quasi-invariant under the whole supergroup. On other hand, for even dimensions, the symmetry breaks into $\mathcal{G} \times SO(1,D-1)$. For example, for $D=4$, the super-group can have no invariant traces, and this is the reason why supersymmetry breaks down. The action in four dimensions must be seen as an effective description, due to, for instance, a quartic fermionic coupling that shows up and prevents the model from being renormalizable \cite{uncon,uncon1,uncon2}. 
\paragraph*{}The paradigm that the procedure still keeps from standard SUSY is that fermion and bosons can be combined into a unique non-trivial representation of a supergroup. The differences already appear in the scenario where the SUSY works. In this proposal, SUSY is an extension of the symmetries of the {\it tangent space}. Since Dirac fermions lie in the $[(\frac{1}{2},0)\oplus(0,\frac{1}{2})]$-representation of Lorentz group, SUSY is implemented as an extension of the tangent space symmetries. This approach allows us to implement SUSY in any manifold,by looking for the symmetries of the tangent bundle. Another difference is found in the field representations \cite{uncon}.
\paragraph*{}A Chern-Simons $AdS_5$ supergravity is a gauge model based on a SUSY extension of the $AdS_5$ gravity. Based on the no-gravitini approach \cite{uncon} and on the structure of $SO(4,2)$ group, we work with a field that is a 1-form gauge connection \cite{ympg}:
\begin{equation}
\hat{\mathcal{A}} = \hat{e}^a J_a + \frac{1}{2} \hat{\omega}^{ab}J_{ab} + \hat{A}^k T_k + (\bar{\psi}^r\hat{\Gamma}Q_r + \bar{Q}^r \hat{\Gamma} \psi_r) + \hat{b} \mathbb{K},
\end{equation}
where the hat stands for 5-dimensional forms; $\hat{\Gamma} = \hat{e}^a \gamma_a$, with $a,0,...,4$; $k=1,...,\mathcal{N}^2-1$ and $r=1,...,\mathcal{N}$. This 1-form has values in the $SU(2,2|\mathcal{N})$ super-algebra, whose bosonic sector is given by $SU(2,2) \otimes SU(\mathcal{N}) \otimes U(1)$, where $SU(2,2) \simeq SO(4,2)$ \cite{sugcs2}.

\paragraph*{}The infinitesimal gauge transformation is given by $\delta \hat{A} = \hat{d} \epsilon + [ \hat{A}, \epsilon\}$, with $\epsilon = \epsilon^a J_a + \frac{1}{2} \epsilon^{ab} J_{ab} + \epsilon^k T_k + \bar{\chi}^rQ_r + \bar{Q}^r \chi_r + \epsilon_b \mathbb{K}$. In components, we have:
\begin{subequations}
\begin{equation}
\delta \hat{e}^a = \hat{d}\epsilon^a + \hat{\omega}^{ab} \epsilon_b + \epsilon^{ab} \hat{e}_b+\frac{1}{2}(\bar{\psi}^r \hat{\Gamma} \gamma^a \chi_r +  \bar{\chi}^r\gamma^a \hat{\Gamma} \psi_r) ~,
\end{equation}
\begin{equation}\delta \hat{\omega}^{ab} = \hat{d}\epsilon^{ab} + \hat{\omega}^{ac} \epsilon_c^{~b} +\hat{\omega}^{bc} \epsilon_c^{~a} + \frac{1}{4}(\bar{\psi}^r \hat{\Gamma} \gamma^{ab} \chi_r +  \bar{\chi}^r\gamma^{ab} \hat{\Gamma} \psi_r)~,
\end{equation}
\begin{equation} \delta \hat{A}^k = \hat{d} \epsilon^k + f^{k}_{~lm}\hat{A}_l \epsilon^m - i (\bar{\psi}^r (\tau^k)_r^{~s}\hat{\Gamma} \chi_s +  \bar{\chi}^r \hat{\Gamma} (\tau^k)_r^{~s}\psi_s)~,
\end{equation}
\begin{equation}\delta( \hat{\Gamma} \psi_r) = \hat{\nabla} \chi_r\end{equation}
\begin{equation}\delta \hat{b} = \hat{d}\epsilon_b + i (\bar{\psi}^r \hat{\Gamma} \chi_r +  \bar{\chi}^r\hat{\Gamma} \psi_r)~,
\end{equation}
\end{subequations}
where  $\hat{\vec{\nabla}} \chi_r = \hat{d}\chi_r +[i(\frac{1}{4}- \frac{1}{\mathcal{N}})\hat{b} + \frac{1}{2}  \hat{e}_a \gamma^a+ \frac{1}{4} \hat{\omega}_{ab} \gamma^{ab}]\chi_r + \hat{A}_k (\tau^k)_r^{~s}\chi_s$. 
\paragraph*{}In the work of Ref.\cite{uncon1}, to ensure that no gravitini appear in the spectrum in a 3D action, the authors show that the dreibein remains invariant under gauge and supersymmetry transformations, but rotates as a vector under the Lorentz subgroup. To do this, we must look for the SUSY transformations. In the fermionic part, we have $\delta( \hat{\Gamma} \psi_r) = \hat{\vec{\nabla}} \chi_r$, where $\chi$ is the local SUSY parameter. Any vector with spinor index can be split into irreducible representations: $1 \otimes 1/2 = 3/2 \oplus 1/2$ of the Lorentz group. So, for $\xi_a^\alpha =( P_{3/2} + P_{1/2} )_b^a\xi_b^\alpha = \phi_a^\alpha + \Psi_a^\alpha$, where $(P_{3/2})_a^b = \delta_a^b - \frac{1}{5}\gamma_a \gamma^b = \delta_a^b - (P_{1/2})_a^b$ are the projectors, $\phi_a^\alpha$ are the 3/2-component and $\Psi_a^\alpha$ is the 1/2-component. Therefore, we have $(P_{3/2})_a^b \gamma_b \psi = 0$, by definition. SUSY transformation yields:
\begin{equation}\label{etransf}
\delta( \hat{\Gamma} \psi_r) = \delta \hat{e}^a \gamma_a \psi_r + \hat{e}^a \gamma_a \delta \psi_r = \hat{\vec{\nabla}} \chi_r.
\end{equation}
\paragraph*{}Applying the $P_{3/2}$-projector to the equation above, we find that \begin{equation}(P_{3/2})_\mu^{~\nu} \hat{\nabla}_\nu \chi_r =0~,\end{equation} which implies that $\hat{\nabla} \chi_r = \hat{e}^a \gamma_a \rho_r$, for an arbitrary spinor $\rho$. This condition guarantees that the symmetry transformations close off-shell without the need of introducing auxiliary fields \cite{uncon1}. Applying $P_{1/2}$-projector to the equation \eqref{etransf} we obtain that, under SUSY, $\delta \psi_r = \rho_r$ and $\delta \hat{e}^a = 0$. The spinor $\rho_r$ obeys the Killing equation; the number of Killing spinors defines the number of unbroken supersymmetries, i.e., supersymmetries respected by the background \cite{uncon1}. For instance, if $\rho_r = 0$, we have $\chi_r = \text{constant}$ (covariantly constant), and we obtain a global SUSY. For a general solution, a Hamiltonian analysis must be carried out to extract the exact solution for the SUSY parameter \cite{uncon2}.

\paragraph*{}The field-strength is given by $\hat{F} = \hat{d} \hat{A} + \frac{1}{2}[\hat{A}, \hat{A}\}$. In components, we have $\hat{F} = \hat{F}^a J_a + \frac{1}{2}\hat{F}^{ab}J_{ab} + \hat{F}^k T_k + \bar{\hat{\Theta}}^r Q_r + \bar{Q}^r \hat{\Theta}_r + F \mathbb{K}$, where:
\begin{subequations}

\begin{equation}
\hat{F}^{a} = \hat{d} \hat{e}^a +  \hat{\omega}^{a}_{~b} \hat{e}^b  + \bar{\psi}^r\hat{\Gamma}  \gamma^a \hat{\Gamma} \psi_r = \hat{D}_{\hat{\omega}} \hat{e}^a + \bar{\psi}^r\hat{\Gamma}  \gamma^a \hat{\Gamma} \psi_r~, 
\end{equation}
\begin{equation}\hat{F}^{ab} = \hat{R}^{ab} + \hat{e}^a \hat{e}^b + \frac{1}{2}\bar{\psi}^r \hat{\Gamma}  \gamma^{ab} \hat{\Gamma} \psi_r ~,
\end{equation}
\begin{equation}
\hat{F}^k = \hat{d} \hat{A}^k +  f^{k}_{~lm}\hat{A}^{l} \hat{A}^m  + \bar{\psi}^r\hat{\Gamma} 
(\tau^k)_r^{~s}\hat{\Gamma} \psi_s~,
\end{equation}

\begin{equation}
\hat{\Theta}_r = (\hat{\nabla})_r^s(\hat{\Gamma}\psi_s) \text{~~,~~} \bar{\hat{\Theta}}^r =  -(\hat{\nabla})^r_s(\bar{\psi}^s\hat{\Gamma})~,
\end{equation}
\begin{equation}
\hat{F} =  \hat{d} \hat{b} + i \bar{\psi}^r \hat{\Gamma} \hat{\Gamma} \psi_r ,
\end{equation}
\end{subequations}
where $\hat{R}^{ab} = \hat{d}\hat{\omega}^{ab}+\hat{\omega}^{ac}\hat{\omega}_c^{~b}$. In the sequel, we shall specifically analyze the SUSY transformations and see how the gravitino sector is suppressed from the model. This Letter follows the outline bolow: In Section II, we displays the salient properties of the model, the action and the associated equations of motion. In Section III, we present our dimensional reduction scheme, Randal-Sundrum's presciption. Next, in the Section IV, we analyze possible non-trivial fermionic solutions and, finally, in Section V, we cast our Concluding Comments. 
\section{5D topological Action }\label{sec3}
\paragraph*{}The topological action we shall be dealing with can be written as a Chern-Simons action in 5 dimensions \cite{sugcs2}:
\begin{equation}
S^{5D} = \int \langle \mathcal{A}  \mathcal{F}  \mathcal{F} - \frac{1}{2} \mathcal{F}\mathcal{A}\mathcal{A}\mathcal{A} + \frac{1}{10}\mathcal{A}\mathcal{A}\mathcal{A}\mathcal{A}\mathcal{A} \rangle ,
\end{equation}
\paragraph*{}where $\langle ... \rangle$ stands for the supertrace. Using these definitions, we can split the components of the action $S^{5D} = S_G + S_{SU(\mathcal{N})} + S_{U(1)} + S_f$, so that:
\begin{equation}
S_G = -\frac{1}{2}\epsilon_{abcde} \int \hat{F}^{ab} \hat{F}^{cd} \hat{e}^e - \frac{1}{2} \hat{F}^{ab} \hat{e}^c\hat{e}^d\hat{e}^e + \frac{1}{10} \hat{e}^a\hat{e}^b\hat{e}^c\hat{e}^d\hat{e}^e~,
\end{equation}

\begin{eqnarray}\nonumber
S_{SU(\mathcal{N})} &=& - \int Tr\Big{[}{\bf \hat{A}} {\bf \hat{F}}  {\bf \hat{F}} - \frac{1}{2} {\bf \hat{A}}{\bf \hat{A}} {\bf \hat{A}} {\bf \hat{F}} + \frac{1}{10}{\bf \hat{A}}{\bf \hat{A}}{\bf \hat{A}}{\bf \hat{A}}{\bf \hat{A}}\Big{]} +\\&& +\frac{i}{2} {\bf \hat{A}} \cdot \bar{\psi}^r\hat{\Gamma} (\hat{\nabla} {\bf \tau} \hat{\nabla})_r^{~s} \hat{\Gamma} \psi_s ~,
\end{eqnarray}

\begin{eqnarray}\nonumber
S_{U(1)} &=& \int (\frac{1}{\mathcal{N}^2} + \frac{1}{4^2}) \hat{b} (\hat{F})^2 + \hat{b}\Big{(}-\frac{1}{4}\hat{F}^{ab}  \hat{F}_{ab} - \frac{1}{4} \hat{F}^a\hat{F}_a+\\\nonumber
&& - \frac{1}{\mathcal{N}}\hat{F}^i \hat{F}_i + \frac{1}{2}(\frac{1}{4}+ \frac{1}{\mathcal{N}})\bar{\psi}^r \hat{\Gamma} (\hat{\nabla}^2)_r^{~s} \hat{\Gamma} \psi_s\Big{)}~,\\
\end{eqnarray}
and
\begin{equation}
S_f = i \int \bar{\psi}^r \hat{\Gamma}\hat{\mathcal{R}}_{ r}^{~s}(\hat{\nabla} \hat{\Gamma} \psi)_s  + c.c.~,
\end{equation}
\paragraph*{}here, \begin{eqnarray}\nonumber
(\hat{\nabla}^2)_r^s &=& \big{[}\frac{1}{4}(\hat{R}^{ab} + \hat{e}^a\hat{e}^b)\gamma_{ab} + \frac{1}{2}\hat{T}^a \gamma_a + i (\frac{1}{4}-\frac{1}{\mathcal{N}})\hat{d}\hat{b}\big{]}\delta_{r}^s + \\&&+\big{[}\hat{d}\hat{A}^k  + f^{kk'k''}\hat{A}^{k'} \hat{A}^{k''}\big{]}(\tau^k)_r^s~,
\end{eqnarray}
\begin{equation}
(\hat{\nabla} \tau^k \hat{\nabla})_r^s = (\hat{\nabla}^2)_r^{s'}(\tau^k)_{s'}^s ,
\end{equation}
 and \begin{equation}\hat{\mathcal{R}}_r^{~s} = \big{[}- \frac{1}{4} \hat{F}^{ab} \gamma_{ab} - \frac{1}{2} \hat{F}^a \gamma_a + \frac{i}{2}(\frac{1}{4}+ \frac{1}{\mathcal{N}})\hat{F}\big{]}\delta_r^{~s} + \hat{F}^i (\tau_i)_r^{~s}~.\end{equation}
 \paragraph*{}It should be noticed that, since $\hat{\mathcal{R}}_r^s \supset \hat{\Gamma} \hat{\Gamma}\delta_r^s$, the fermionic part of $S_f$ generates a Dirac-like action for the fermions ($S_f \supset \int d^5 x \bar{\psi}^r \slashed{D} \psi_r$). Notice that the bosonic part is almost the same in comparison with the usual 5D AdS-SUGRA action \cite{sugcs2}. The important difference lies in the fermionic sector which we will analyze in the Section IV.
\subsection*{Inspecting the Field equations} 
\paragraph*{}The  5D-CS action transforms under a gauge transformation as $\delta S^{5D} = \int \langle \mathcal{F} \mathcal{F} \delta \mathcal{A} \rangle$. We can check that, due to this result, one can readily find the field equations in terms of component fields; they are given by: 
\begin{equation}
\delta \hat{e}^a \rightarrow -\frac{1}{2}\varepsilon_{a b c d e} \hat{F}^{b c} \hat{F}^{d e} -\frac{1}{4} \hat{F}_{b} \hat{F} - \frac{i}{2} \bar{\hat{\Theta}}^r \gamma_a \hat{\Theta}_r =0~,
\end{equation}

\begin{equation}
\delta \hat{\omega}^{ab} \rightarrow -\frac{1}{2}\varepsilon_{a b c d e} \hat{F}^{c d} \hat{F}^e -\frac{1}{4} \hat{F}_{ab} \hat{F} - \frac{i}{2} \bar{\hat{\Theta}}^r \gamma_{ab} \hat{\Theta}_r =0~,
\end{equation}

\begin{eqnarray}\nonumber
\delta \hat{b} &\rightarrow& -\frac{1}{4} \hat{F}^{ab} \hat{F}_{ab} -\frac{1}{4} \hat{F}^{a} \hat{F}_a -\frac{1}{\mathcal{N}}\hat{F}^i \hat{F}_i +(\frac{1}{\mathcal{N}^2}-\frac{1}{4^2}
)(\hat{F})^2+\\&& -\frac{1}{2}(\frac{1}{4}- \frac{1}{\mathcal{N}}) \bar{\hat{\Theta}}^r \hat{\Theta}_r=0~,
\end{eqnarray}

\begin{equation}
\delta \hat{A}^{i} \rightarrow  f^{ikj}\hat{F}^{j} \hat{F}^k +\frac{1}{\mathcal{N}} \hat{F}_i \hat{F} + \frac{i}{2} \bar{\hat{\Theta}}^r (\tau^i)_r^s \hat{\Theta}_s = 0~,
\end{equation}

\begin{equation}
\mathcal{R}_r^{~s} \hat{\Theta}_s  = 0~.
\end{equation}
\paragraph*{}It can be checked that $\mathcal{F}= 0$ is a solution to the field equation. Let us analyze this solution. In components, we have:
\begin{subequations}

\begin{equation}
\hat{F}^{a} = 0 \rightarrow  \hat{T}^a=\hat{D}_{\hat{\omega}} \hat{e}^a =- \bar{\psi}^r\hat{\Gamma}  \gamma^a \hat{\Gamma} \psi_r 
\end{equation}
\begin{equation}
\hat{F}^{ab} = 0 \rightarrow \hat{R}^{ab} + \hat{e}^a \hat{e}^b = - \frac{1}{2}\bar{\psi}^r \hat{\Gamma}  \gamma^{ab} \hat{\Gamma} \psi_r 
\end{equation}
\begin{equation}
\hat{F}^k = 0 \rightarrow \hat{d} \hat{A}^k +  f^{k}_{~lm}\hat{A}^{l} \hat{A}^m  =- \bar{\psi}^r\hat{\Gamma} 
(\tau^k)_r^{~s}\hat{\Gamma} \psi_s
\end{equation}

\begin{equation}\hat{F} = 0 \rightarrow \hat{d} \hat{b} =- i \bar{\psi}^r \hat{\Gamma} \hat{\Gamma} \psi_r.
\end{equation}
and 
\begin{equation}
\hat{\Theta}_s = 0
\end{equation}
\end{subequations}

\paragraph*{}The first four equations above are not but the equation of motion for the bosonic fields. In the section \ref{secf} we analize the fermionic component $\hat{\Theta}_s = 0$.

\section{On the Dimensional Reduction}

\paragraph*{}We have that the index $a= 0,...,4= I,4$, where the index $I$ refers to the $SO(1,3)$ Minkowski group. So, the fields can be split into two pieces \cite{chams,neves}.
\begin{equation}
\hat{\omega}^{ab} = \{ \hat{\omega}^{IJ}, \lambda \hat{b}^I\} ~~,~~ \hat{e}^a = \{\hat{e}^I , \hat{e}^4\}~.
 \end{equation}
\paragraph*{}Besides that, we are also interested in considering the action in a 4-dimensional version; so, we must split the coordinates as $x^\alpha = (x^\mu, \chi)$ and the 1-forms can be written it as follows below:
\begin{equation}\label{dimred1}
\hat{\omega}^{IJ} = \omega^{IJ} + \omega^{IJ}_\chi d\chi ~~;~~  \hat{b}^I = {b}^I + {b}^I_\chi d\chi  
\end{equation}
\begin{equation}\label{dimred2}
\hat{e}^I = e^I + e^I_\chi d\chi ~~;~~ \hat{e}^4 = e^4 + e^4_\chi d\chi 
\end{equation}
\begin{equation}\label{dimred3}
\hat{b} = b + b_\chi d \chi ~~; ~~ \hat{A}^k = A^k + A^k_\chi d \chi ~.
\end{equation}
\paragraph*{} Since the 5 D gamma-matrices can be split as $\gamma^a = (\gamma^I, \gamma_5)$, we then have $\hat{\Gamma} = \Gamma_\mu dx^\mu + \Gamma_\chi d\chi$, where:
\begin{equation}
\Gamma = \gamma^I e_I + \gamma_5 e^4 ~~;~~ \Gamma_\chi = \Big{[}\gamma^I (e_I)\chi + \gamma_5 e^4_\chi \Big{]} d \chi ~.
\end{equation}

\paragraph*{}As we can see, the equation $\mathcal{F}=0$ is a solution (but, non-unique) to the field equations for the topological action. Therefore, we can analyze this solution in terms of the reduced components (see \ref{app}).

\subsection*{Randall-Sundrum Dimensional Reduction}\label{RSR}

\paragraph*{}A Randall-Sundrum-like ansatz (RS) is proposed with the assumption that the geometry of 5D space-time has the following structure \cite{rand0,rand,rand1}:
\begin{equation}\label{metric4d}
ds^2_{5D} = e^{-2 \sigma(\chi)} g_{\mu \nu}(x) dx^\mu dx^\nu + G(\chi)^2 d\chi^2~. 
\end{equation}
We can translate \eqref{metric4d} in terms of the following fünfbein:
\begin{equation}\label{ansRS} \hat{e}^{a}_\alpha=\begin{bmatrix}
h^{I}_{~\mu}(x) e^{-\sigma(\chi)} & 0\\
0 & G(\chi)\\
\end{bmatrix},
\end{equation}
where $\sigma$ is called conformal function.  The 4D metric can be written as $g_{\mu \nu}(x) = \eta_{IJ} h^{I}_{~\mu} h^{J}_{~\nu}$, with $\eta^{IJ}$ = diag(1,-1,-1,-1). Going further, the inverse of the fünfbein is given by: 
\begin{equation}\hat{E}_{a}^\alpha=\begin{bmatrix}
(h^{-1}(x) )_{I}^{~\mu}e^{\sigma(\chi)} & 0\\
0 & \frac{1}{G(\chi)}\\
\end{bmatrix},
\end{equation}
where we assume that $h^I_{~\mu}$ has an inverse, i.e., $h^I_{~\mu} (h^{-1})^{J\mu} = \eta^{IJ}$. This opens up the opportunity to define the inverse of the vielbein (in 4D). We have that $e^I = e^{-\sigma}h^I_\mu dx^\mu = e^I_\mu dx^\mu$, which implies $e^I_\mu e_{\nu I} = g_{\mu \nu}$. Therefore, we may write $E^\mu_I = e^{\sigma}(h^{-1})^\mu_{~I}$ so that $e^I_\mu E^{\mu}_J = \delta^{I}_{J}$. We can define a operation ``$\rfloor$" in 4D, and we can present the identity above as $E^{\mu}_J e^I_\mu  = (E_J \rfloor e^I) = \delta^{I}_{J}$. We shall use this operation from now on, and the 4D character is implicit in the forms without $`` ~\hat{~}~ "$. 
 
\paragraph*{} From $F^a = 0$, we can directly see that $\omega_\chi^{IJ} = - G \bar{\psi}^r\gamma^{IJ} \gamma_5\psi_r$, $\lambda b^I =- \bar{\psi}^r \Gamma \gamma_5 \gamma^I \psi_r$, $\lambda b_\chi^I= - G \bar{\psi}^r \gamma^I \psi_r= G t^I$ and $\omega^{IJ} = \Omega^{IJ}(e) + 6 i e^M\epsilon_M^{~~IJK}  \bar{\psi}^s \gamma_{K} \gamma_5 \psi_s$ (See Appendix \eqref{app}), and, due to \eqref{ansRS},  $\Gamma = e^I \gamma_I$ .

\section{Fermionic solutions}\label{secf}
With the dimensional reduction ansatz present above, we can now initiate seeking a solution for the fermionic fields $\psi_r$. We have:
\begin{equation}
\hat{\Theta}_r = (\hat{\nabla})_r^s(\hat{\Gamma}\psi_s)=0 \text{~~,~~} \bar{\hat{\Theta}}^r =  -(\hat{\nabla})_r^s(\bar{\psi}^s\hat{\Gamma})=0~;
\end{equation}
here, the covariant derivative can be written as $\hat{\nabla} = (\nabla, \nabla_\chi d \chi)$.
Using the results from $F^a = 0$ and the Randall-Sundrum ansatz the $\hat{\nabla}$-operator must have the following structure:
\begin{eqnarray}\nonumber
\nabla_r^s &=&\Big{(} d +i(\frac{1}{4}- \frac{1}{\mathcal{N}})b + \frac{1}{2}  e_I \gamma^I  + \frac{1}{4} \Omega_{IJ} \gamma^{IJ} -3 (e_I  \bar{\psi}^s \gamma_{J} \gamma_5 \psi_s)\gamma^{IJ}\gamma_5 +  \\
&& -\frac{1}{2}  ( \bar{\psi}^r \Gamma \gamma_5 \gamma^I \psi_r) \gamma_5\gamma_{I}\Big{)}\delta_r^s +A_k (\tau^k)_r^{~s}~,
\end{eqnarray}
and 
\begin{eqnarray}\nonumber
(\nabla_\chi)^s_r &=&\Big{(} \partial_\chi +i(\frac{1}{4}- \frac{1}{\mathcal{N}})b_\chi  + \frac{1}{2} G(\chi) \gamma_5 +\\\nonumber &-&  \frac{G}{4} (  \bar{\psi}^r\gamma_{IJ} \gamma_5\psi_r) \gamma^{IJ} -\frac{G}{2} (  \bar{\psi}^r \gamma^I \psi_r) \gamma_5 \gamma_{I}\Big{)}\delta_r^s + (A_k)_\chi (\tau^k)_r^{~s}~.\\ 
\end{eqnarray}

\paragraph*{}Now, $\hat{\Theta}_{\mu \nu} = 0$ gives us (ignoring the $SU(\mathcal{N})$ indexes temporarily):
\begin{eqnarray}\nonumber
\nabla (\Gamma \psi) &=& \Big{[}(\nabla_\mu e_\nu^I)\gamma^I\psi +  e_\nu^I \nabla_\mu \gamma_I \psi\Big{]}dx^\mu dx^\nu = \\
&& = \Big{[}e_\nu^I \nabla_\mu \gamma_I \psi + T_{\mu \nu}^{~~I}\gamma_I \psi \Big{]}dx^\mu dx^\nu = 0 ~.
\end{eqnarray}
\paragraph*{}Upon contraction of the equation above with $E^\mu_b E^\nu_c$, we arrive at:
\begin{eqnarray}\nonumber
&&\Big{(}E_J E_K \rfloor \nabla (\Gamma \psi) \Big{)}= E^{[\mu}_J E^{\nu]}_K (e_\nu^I \nabla_\mu \gamma_I \psi + T_{\mu \nu}^{~~I}\gamma_I \psi ) = \\
&& \delta^I_{[K} E^\nu_{J]} \nabla_\nu \gamma_I \psi + T_{JK}^{~~I}\gamma_I \psi = \nabla_{[J} \gamma_{K]} \psi - (\bar{\psi} \gamma_{[J}\gamma^I\gamma_{K]} \psi) \gamma_I \psi = 0],
\end{eqnarray}
\paragraph*{}where $\nabla_I = E_I^\mu \nabla_\mu$. Using that $\nabla \gamma_I = \omega^{~J}_I \gamma_J$ and, contracting with $\gamma^J \gamma^K$, we get:

\begin{equation}\label{res1}
\slashed{\nabla} \psi  + \frac{1}{6}\Omega^I_{~JK} \gamma^J \gamma^K \gamma_I \psi = 0~,
\end{equation}
\paragraph*{}where $\Omega^I_{~JK} = E^{\mu I} \Omega_{\mu JK}$, $\slashed{\nabla} = \slashed D + 2 + \frac{1}{2}\Omega_K^{~IJ} \gamma^K \gamma_I \gamma_J  -18 \slashed a \gamma_5 - 2 P \gamma_5 + \frac{1}{4}c^{IJ} \gamma_{IJ}\gamma_5$, $D = d +i(\frac{1}{4}- \frac{1}{\mathcal{N}})b+ A \cdot \tau$, $a^I = \bar{\psi} \gamma^I \gamma_5 \psi $, $P = \bar{\psi} \gamma_5 \psi  $ and $c^{IJ} =  \bar{\psi}\gamma^{IJ} \gamma_5 \psi$. The equation is part of the solution. Going further, the $\hat{\Theta}_{\mu 4}=0$ component reads as below:
\begin{equation}
(\nabla_\chi) ((\gamma^I e_I + \gamma_5 e^4) \psi) - \nabla (\gamma^I (e_I)_\chi + \gamma_5 e^4_\chi)\psi = 0
\end{equation}

\paragraph*{}Using the RS ansatz the equation simplifies to to acquire the form 
\begin{equation}\label{res2}
\slashed{\nabla}(\gamma_5 \psi) - \Delta \psi = 0 ~,
\end{equation}
\paragraph*{}where $\Delta = G^{-1}\gamma^J \Big{(}E_J \rfloor \nabla_\chi( e_I \gamma^I \psi)\Big{)} = (4  G^{-1}D_\chi - 2 \gamma_5 - 16 P  +2 d^I\gamma_I \gamma_5 )\psi$, with $D_\chi = \partial_\chi +i(\frac{1}{4}- \frac{1}{\mathcal{N}})b_\chi+ A_\chi \cdot \tau$, $G^{-1}\sigma' = 4 P$ and $d^I =  \bar{\psi} \gamma^I \psi$. From Eqs. \eqref{res1} and \eqref{res2} we can reach a differential equation for each chiral component of $\psi$. Using the definitions $\psi_L = \frac{1-\gamma_5}{2} \psi$ and $\psi_R = \frac{1+\gamma_5}{2} \psi$ (the splitting the Dirac matrices in the left and right components), redefining the physical fermionic field as $\psi|_{ph} = \sqrt \mu \psi$ and the physical vielbein $e^a|_{ph} = \mu e^a$, where $\mu$ is a parameter with dimension $[\mu] = \text{mass}^1$, we find the following equations:
\begin{equation}\label{Res1}
\sigma^I (D_I + \frac{5i}{6} \Omega_I - \frac{18}{\mu^2} a_I) \psi_R + (2 \mu + \frac{2}{\mu^2} P - \frac{1}{4 \mu^2}c^{IJ}\sigma_{IJ})\psi_L = 0~,
\end{equation}
 
\begin{equation}\label{Res2}
\bar{\sigma}^I (D_I - \frac{5i}{6} \Omega_I + \frac{18}{\mu^2} a_I) \psi_L + (2 \mu - \frac{2}{\mu^2}P + \frac{1}{4\mu^2}c^{IJ}\bar{\sigma}_{IJ})\psi_R = 0~,
\end{equation}

and 

\begin{equation}\label{Res3}
(4 G^{-1} D_\chi + 2 \mu - \frac{16}{\mu^2} P )\psi_L - \sigma^I ( \frac{i}{6} \Omega_I - \frac{2}{\mu^2} d_I) \psi_R= 0~,
\end{equation} 

\begin{equation}\label{Res4}
(4  G^{-1} D_\chi - 2 \mu - \frac{16}{\mu^2} P )\psi_R - \bar{\sigma}^I ( \frac{i}{6} \Omega_I + \frac{2}{\mu^2} d_I) \psi_L= 0~.
\end{equation} 
\paragraph*{}Here we redefine $\Omega^I$ and $G$ in way to possess the correct dimensions $[\Omega^I] = \text{mass}^1$ and $[G] = \text{mass}^0$. The Eqs. \eqref{Res1} and \eqref{Res2} are the Dirac-like equations of the fermionic fields on the brane with mass $M=2 \mu$. Moreover, the equations \eqref{Res3} and \eqref{Res4} are the cubic field equations from which we can fix the dependence on the extra dimension $\chi$. For instance, in the strong limit of $\mu$ parameter, assuming $b_\chi=0$ and $A_\chi=0$, and for flat brane limit ($e_\mu^I = e^{-\sigma}\delta_\mu^I$ ), we are able to ignore the self-interaction terms, and we can write the fermionic fields as $\psi_L = \alpha(\chi)\psi_R(x) $ and $\psi_R = \beta(\chi)\psi_R(x)$, where $x$ represents the coordinates inside the 4-brane. This are the zero mode solutions of the RS context. Therefore, we find the following solutions for $\alpha$ and $\beta$:
\begin{equation}
\alpha(\chi) = \alpha_0 e^{\frac{1}{2}\int_0^\chi d\chi' G(\chi') } ~~,~~ \beta(\chi) = \beta_0 e^{-\frac{1}{2}\int_0^\chi d\chi' G(\chi') }~,
\end{equation}
\paragraph*{}where $\alpha_0$ and $\beta_0$ are constants. In this limit, we recover the well-known result which points to the problem of the simultaneous localization of the chiralities in the brane. Besides that, the ansatz are strongly dependent on the linear approximation and the flat brane limit.  
\paragraph*{}Going further, we can contemplate the case $b_\chi = \Phi(\chi)$. In this situation, the solutions for $\alpha$ and $\beta$ can still be found and they are the following:
\begin{equation}
\alpha(\chi) = \alpha_0 e^{\frac{1}{2}\int_0^\chi d\chi' \big{[}G(\chi') + 2 i(\frac{1}{4}- \frac{1}{\mathcal{N}})\Phi(\chi')/\mu\big{]} } ~~;~~ \beta(\chi) = \beta_0 e^{-\frac{1}{2}\int_0^\chi d\chi'\big{[}G(\chi') - 2 i(\frac{1}{4}- \frac{1}{\mathcal{N}})\Phi(\chi')/\mu \big{]}} 
\end{equation}
\paragraph*{}As we can see, the $\Phi$-field can modify the localization profile by a complex phase. A similar consideration could be made considering the $A_\chi$-field. Although, the $SU(\mathcal{N})$ structure requires more care. If $A_\chi = A_{\chi}(\chi)$ is considered, it generates an $SU(\mathcal{N})$ phase, analogous to the $\Phi$-phase. Despite the absence of an imaginary $i$ multiplying $A_\chi$, the contribution maintains as a complex phase, due to the anti-hermitian character of the generators of $SU(\mathcal{N})$ adopted here ( see Ref.\cite{5dCS}).

\paragraph*{}If we analyze the full set of equations, a non-linear system appears as a challenge for the comprehension of the spinor-localization in the branes. We shall focus on that and we are going to report on this issue in a forthcoming paper.

\section{Concluding Comments}

\paragraph*{}As shown previously, the fermionic equation of motion we have worked out is a non-linear equation, and the fermion field displays some interesting couplings. First of all, we find that the fermion acquires mass $M = 2 \mu$, with $\mu$ a parameter that appears naturally in the formalism. As we have seen, the $\chi$-component of the bosonic field $b$ does not interfere in the localization scheme; the same happens with the $\chi$ component of the bosonic field $A$, at least in the strong mass limit. As it can be checked \cite{loc1}, is possible to localize both chiralities in the brane, by virtue of the presence of torsion. A careful analysis should be implemented to confirm this hint. 
\paragraph*{}Another question remains to be clarified in a forthcoming publication. In ordinary Randall-Sundrum scenarios, only one of the fermion chiralities ($\psi_L$ or $\psi_R$) can be localized in the brane. Does this conclusion persist if quartic self-couplings are considered? May the non-linearity induce some new behavior of the localization scheme (i.e., topological solutions)? Is important to highlight that this non-linearity is present in other models of gravity with fermionic degrees of freedom. In our case, local SUSY unavoidably carries an intrinsic torsion from the new fermionic degrees of freedom.  
\paragraph*{}By means of Fierzings, we can generate a Nambu-Jona-Lasinio cubic term in the field equations, and a dynamical symmetry breaking might take place. This is expected in theories with torsion, but the particularity here is that the coupling constant could be $\chi$-dependent. This property may induce different masses, depending on the brane localization and, together with the mass $M$, it could yield a ``see-saw-like" mechanism, involving a light and a heavy chirality.   
\paragraph*{}

\section*{Acknowledgments}
Thanks are due to Prof.  J.A. Helayel-Neto for discussions and suggestions on the original manuscript. This work is funded by the Brazilian National Council for Scientific and Technological Development (CNPq-Brasil). 
\section*{Appendix}\label{app}
\paragraph*{}The representation of the generators is given in terms of $(4 + \mathcal{N}) \times (4 + \mathcal{N})$-supermatrices \cite{sugcs,sugcs1}:\\
\\{
\begin{eqnarray}\nonumber
&&J_{ab}=\begin{bmatrix}
\frac{1}{2} (\gamma_{ab})^\alpha_{~\beta} & 0\\
0 & 0\\
\end{bmatrix} ~~,~~ J_{a}=\begin{bmatrix}
 (\gamma_{a})^\alpha_{~\beta} & 0\\
0 & 0\\
\end{bmatrix} ~~,~~ \\\nonumber
&& T_k=\begin{bmatrix}
0 & 0\\
0 & (\tau^k)_r^{~s}\\
\end{bmatrix} ~~,~~ 
Q^\alpha_s=\begin{bmatrix}
0 & 0\\
- \delta^r_s \delta^\alpha_\beta & 0\\
\end{bmatrix} ~~,~~ \\
&&\bar{Q}_\alpha^s=\begin{bmatrix}
0 & \delta^r_s \delta^\alpha_\beta\\
0 & 0\\
\end{bmatrix} ~~,~~ \mathbb{K}=\begin{bmatrix}
\frac{i}{4}\delta^\alpha_\beta & 0\\
0 & \frac{1}{\mathcal{N}}\delta^{~s}_r\\
\end{bmatrix}~.
\end{eqnarray}
}\\
\paragraph*{}From that, the following algebra can be written:
\begin{eqnarray}\nonumber
&&[J^{ab}, J^{cd}] = \eta^{ad}J^{bc} - \eta^{ac}J^{bd} + \eta^{bc}J^{ad} -\eta^{bd}J^{ac}  ~~ , ~~\\\nonumber
&& [J^a, J^b] = s^2 J^{ab} ~~,~~ [J^a, J^{bc}] =\eta^{ab} J^c - \eta^{ac} J^b~~,~~\\\nonumber
&&[J^a, Q_s]= -\frac{s}{2}\gamma^a Q_s ~~,~~ [J^a, \bar{Q}^s]=  \frac{s}{2}\bar{Q}^s\gamma^a ~~,~~\\\nonumber && [J^{ab}, Q_s] = -\frac{1}{2} \gamma^{ab} Q_s ~~,~~[J^{ab}, \bar{Q}^s] = \frac{1}{2} \bar{Q}^s\gamma^{ab} ~~,~~\\\nonumber&& [\mathbb{K}, Q_s] = -i(\frac{1}{4}- \frac{1}{\mathcal{N}}) Q_s ~~,~~ [\mathbb{K}, \bar{Q}^s] =  i(\frac{1}{4}- \frac{1}{\mathcal{N}}) \bar{Q}^s ~~,~~ \\\nonumber &&[T^k, Q_s] = (\tau^k)_s^r Q_r~~,~~ [T^k, \bar{Q}^s] = -(\tau^k)^s_r \bar{Q}^r\\\nonumber&& \{Q_s, \bar{Q}^r \} = - \frac{i}{2} \delta^r_s \gamma^a J_a - \frac{1}{4}\delta^r_s\gamma^{ab} J_{ab} + i \delta^r_s \mathbb{K} + (\tau^k)_s^r T_k ~.\\
\end{eqnarray}
\paragraph*{}All the other relations vanish. The only non-vanishing supertraces are:
\begin{equation}\nonumber
\langle J_a J_{b c} J_{d e} \rangle = -\frac{1}{2} \varepsilon_{abcde} ~~,~~  \langle T^i T^j T^k \rangle  = - f^{ijk}
\end{equation}
\begin{equation}\nonumber
\langle \mathbb{K} J_{a b} J_{c d} \rangle = -\frac{1}{4}\eta_{ab,cd} ~~,~~  \langle \mathbb{K} T^i T^j \rangle  = - \frac{1}{\mathcal{N}} \delta^{ij}
\end{equation}
\begin{equation}\nonumber
\langle \mathbb{K} J_{a } J_{b} \rangle = -\frac{1}{4}\eta_{a b} ~~,~~  \langle \mathbb{K}\mathbb{K}\mathbb{K} \rangle = \frac{1}{\mathcal{N}^2} + \frac{1}{4^2} 
\end{equation}
\begin{equation}\nonumber
\langle \bar{Q}^\alpha_r J_{a b} Q_\beta^s \rangle = -\frac{i}{4}(\Gamma_{ab})^\alpha_\beta  \delta^s_r ~~,~~  \langle \bar{Q}^\alpha_r T^i Q_\beta^s \rangle  = -\frac{i}{2}\delta^{\alpha}_\beta (\tau^i)^s_r
\end{equation}
\begin{equation}\nonumber
\langle \bar{Q}^\alpha_r J_{a} Q_\beta^s \rangle = -\frac{i}{2}(\Gamma_{a})^\alpha_\beta  \delta^s_r ~~,~~  \langle \bar{Q}^\alpha_r \mathbb{K} Q_\beta^s \rangle  = -\frac{1}{2}(\frac{1}{4} + \frac{1}{\mathcal{N}})\delta^{\alpha}_\beta \delta^s_r
\end{equation}
\paragraph*{}In the dimensionally reduced scenario, we have that the covariant derivative can be written as $\hat{\nabla} = (\nabla, \nabla_\chi d \chi)$, where:
\begin{eqnarray}\nonumber
\nabla_r^s &=&\Big{(} d +i(\frac{1}{4}- \frac{1}{\mathcal{N}})b + \frac{1}{2}  e_I \gamma^I + \frac{1}{2} e^4 \gamma_5 + \frac{1}{4} \omega_{IJ} \gamma^{IJ} +\\&+&  \frac{\lambda}{2}  b_I \gamma_5\gamma^{I}\Big{)}\delta_r^s +A_k (\tau^k)_r^{~s}~,
\end{eqnarray}
\begin{eqnarray}\nonumber
(\nabla_\chi)^s_r &=&\Big{(} \partial_\chi +i(\frac{1}{4}- \frac{1}{\mathcal{N}})b_\chi + \frac{1}{2}  (e_I)_\chi \gamma^I + \frac{1}{2} e^4_\chi \gamma_5 +\\\nonumber &+&  \frac{1}{4} (\omega_{IJ})_\chi \gamma^{IJ} +\frac{1}{2} \lambda (b_I)_\chi \gamma_5 \gamma^{I}\Big{)}\delta_r^s + (A_k)_\chi (\tau^k)_r^{~s}~,\\ 
\end{eqnarray}
\paragraph*{}The above equations, components of $F^a = 0$, are useful to the fermionic equations of motion:
\begin{subequations}
 \begin{equation}\label{torsion}
 d e^I + \omega^{I}_{~J}e^J  = T^I =  - \bar{\psi}^r \Gamma\gamma^I\Gamma\psi_r~,
 \end{equation}
 \begin{equation} 
-\sigma' e^I + \omega_\chi^{IJ} e_J = - G \bar{\psi}^r \Gamma\gamma^I \gamma_5  \psi_r~, 
\end{equation}
\begin{equation}
\lambda b^I e_I =- \bar{\psi}^r \Gamma\gamma_5 \gamma^I \psi_r e_I~,
\end{equation}
\begin{equation} 
\lambda b_\chi^I e_I= - G \bar{\psi}^r \gamma^I \psi_r e_I ~,
\end{equation}
\end{subequations}
\paragraph*{}which gives us $\omega_\chi^{IJ} = - G \bar{\psi}^r\gamma^{IJ} \gamma_5\psi_r$, $\lambda b^I =- \bar{\psi}^r \Gamma \gamma_5 \gamma^I \psi_r$, $\lambda b_\chi^I= - G \bar{\psi}^r \gamma^I \psi_r$ and $\sigma'(\chi) = 4 G \bar{\psi}^r \gamma_5 \psi_r$. From Eq.\eqref{torsion}, we can write the spin connection $\omega^{IJ}$ as $\omega^{IJ} = \Omega^{IJ}(e) + 6 i e^M\epsilon_M^{~~IJK}  \bar{\psi}^s \gamma_{K} \gamma_5 \psi_s$, where $\Omega^{IJ} =  \varepsilon^{IJKL}e_K \Omega_L$ and $\Omega^J = - \varepsilon^{MN~J}_{~~~~I}(E_N E_M \rfloor d e^I)$

\end{document}